\documentclass[prd,eqsecnum,twocolumn, 
showpacs,nofootinbib,superscriptaddress]{revtex4}
\usepackage{bm}
\usepackage{stmaryrd}
\usepackage{amsmath}
\usepackage{amssymb}
\usepackage{graphicx}
\usepackage{textcomp}
\usepackage{calrsfs}
\usepackage{yfonts}

\newcommand{\nn}{\nonumber}

\newcommand{\be}{\begin{equation}}
\newcommand{\ee}{\end{equation}}
\newcommand{\ben}{\begin{equation*}}
\newcommand{\een}{\end{equation*}}
\newcommand{\bea}{\begin{eqnarray}}
\newcommand{\eea}{\end{eqnarray}}

\newcommand{\vect}[1]{\bm{#1}}

\newcommand{\sprod}{\!\cdot\!}

\newcommand{\bdalpha}{{\boldsymbol \alpha}}

\newcommand{\uvx}{\hat{\mathbf{x}}}
\newcommand{\uvy}{\hat{\mathbf{y}}}
\newcommand{\uvz}{\hat{\mathbf{z}}}

\DeclareMathOperator{\Tr}{Tr}
\DeclareMathOperator{\tr}{tr}

\begin{document}

\title{Casimir-Polder repulsion: Three-body effects}

\date{\today}

\author{Kimball A. Milton}\email{milton@nhn.ou.edu}
\affiliation{Homer L. Dodge Department of
Physics and Astronomy, University of Oklahoma, Norman, OK 73019-2061, USA}
\author{E. K. Abalo}\email{ekabalo@gmail.com}
\affiliation{Homer L. Dodge Department of
Physics and Astronomy, University of Oklahoma, Norman, OK 73019-2061, USA}
\author{Prachi Parashar}\email{prachi@nhn.ou.edu}
\affiliation{Homer L. Dodge Department of
Physics and Astronomy, University of Oklahoma, Norman, OK 73019-2061, USA}
\affiliation{Departmeht of Physics, Southern Illinois University--Carbondale,
Carbondale, IL 62901 USA}

\author{Nima Pourtolami}\email{nima.pourtolami@gmail.com}
\affiliation{Department of Physics,
Concordia University,
7141 Sherbrooke St.~W., SP-367.7, Montreal, Quebec, Canada H4B 1R6}
\author{Iver Brevik}\email{iver.h.brevik@ntnu.no}
\affiliation{Department of Energy and Process Engineering,
Norwegian University of Science and Technology, N-7491 Trondheim, Norway}
\author{Simen \AA. Ellingsen}\email{simen.a.ellingsen@ntnu.no}
\affiliation{Department of Energy and Process Engineering,
Norwegian University of Science and Technology, N-7491 Trondheim, Norway}
\author{Stefan Yoshi Buhmann}\email{stefan.buhmann@physik.uni-freiburg.de}
\affiliation{Physikalisches Institut, Albert-Ludwigs-Universit{\"a}t Freiburg, 
Hermann-Herder-Str.~3, 79104 Freiburg, Germany and
Freiburg Institute for Advanced Studies, Albert-Ludwigs-Universit{\"a}t 
Freiburg, Albertstr.~19, 79104 Freiburg, Germany}
\author{Stefan Scheel}\email{stefan.scheel@uni-rostock.de}
\affiliation{Institut f\"ur Physik, 
Universit\"at Rostock, Universit\"atsplatz 3,
D-18051 Rostock, Germany}

\begin{abstract}  In this paper we study an archetypical scenario in which
repulsive Casimir-Polder forces between an atom or molecule and two
macroscopic bodies can be achieved.  This is an extension of previous 
studies of the interaction between a polarizable 
atom and a wedge, in which repulsion occurs if the atom
is sufficiently anisotropic and close enough to the symmetry plane
of the wedge. A similar repulsion occurs if such an atom passes a thin
cylinder or a wire.  
An obvious extension is to compute the interaction between such an
atom and two facing wedges, which includes as a special case the
interaction of an atom with a conducting screen possessing a slit, or between
two parallel wires.  To this end we further extend the electromagnetic 
multiple-scattering formalism
for three-body interactions.  To test this machinery we
reinvestigate the interaction of a polarizable atom  between two parallel
conducting plates.  In that case, three-body effects are shown to be small,
and are dominated by three- and four-scattering terms.  The atom-wedge
calculation is illustrated by an analogous scalar situation, described
in the Appendix.
The wedge-wedge-atom geometry is difficult to analyze because this is a 
scale-free problem.  But it is not so hard to 
investigate the three-body corrections to the interaction between 
an anisotropic atom or nanoparticle and a pair of parallel conducting
cylinders, and show that the three-body effects are very small and do not 
affect the Casimir-Polder repulsion at large distances between the cylinders.
Finally, we consider whether such highly anisotropic atoms needed for repulsion
are practically realizable.  Since this appears rather difficult to
accomplish, it may be more feasible to observe such effects with highly
anisotropic nanoparticles.

\end{abstract}

\pacs{42.50.Lc, 32.10.Dk, 12.20.-m, 03.70.+k}
\maketitle

\section{Introduction}

It is quite remarkable that after nearly seven decades, interest in
the so-called Casimir effect \cite{Casimir:1948dh} remains so high.
There have been many theoretical and experimental developments in the
last few years.   For a review of the
status of quantum vacuum energy phenomena in
general, the reader is referred to the volume edited by Dalvit 
et al.~\cite{dalvitvol}.

One of the hottest topics in the field is the subject of repulsive
Casimir effects.  This could have a major impact in nanotechnology,
where at distances well below 1 $\mu$m, 
Casimir forces can play a dominating role.
Repulsion can occur between electric and magnetic bodies, between
electric bodies separated by a medium with an intermediate value of
permittivity, or purely due to the geometry of the two bodies.  
The geometric repulsion that was demonstrated in Ref.~\cite{Levin:2010zz}
was based on numerical methods.  A great deal of effort has been given to
understanding the underlying analytical structure of repulsive effects
\cite{Milton:2011ni,Milton:2011ed,Shajesh:2011np}.
A brief review, with references, is provided in
Ref.~\cite{Milton:2012ti}.  For further work on repulsion
see Refs.~\cite{bostrom12,bostrom13,dou14}.

In this paper, we will further develop the multiple-scattering approach
to include three-body effects, which was introduced in the scalar
context in Refs.~\cite{Emig:2007me,Schaden:2010wv,shajesh},
in particular in the context of Casimir
repulsion.  The electromagnetic formulation
 is in turn based on Green's dyadics, which have a long
history. The Green's dyadic approach to computing the Casimir effect was first
proposed in Refs.~\cite{Schwinger:1977pa,Milton:1978sf}. This was a
tensorial generalization of the scalar Green's function variational
approach Schwinger had given a few years earlier \cite{schwingerlmp}.
All of this was in the direct line of evolution to what is now
referred to as the multiple scattering method.

Although it is well-appreciated that Casimir forces are not additive,
most work on such interactions has concentrated on forces between two bodies.
But the multiple-scattering formalism is easily generalized to include
three-body interactions.
In Sec.~\ref{sec:3body} of this paper
we develop the three-body formalism for the electromagnetic case.  
This is relevant to computing the force between
a polarizable atom and two co-planar half planes, forming a slit, for
example.  As a simple illustration of the formalism, we re-examine, in
Sec.~\ref{sec:parallel}, the 
interaction between an atom and two parallel conducting plates, first
considered by Barton \cite{barton}.  Naturally, the two-body forces dominate
when the atom is near one or the other of the plates, and three-body effects
become significant at the several percent level only when the atom is roughly
equidistant from the two plates, but then the force is quite small.
In Ref.~\cite{Milton:2013if} we examined the nonmonotonicity that can arise 
when two polarizable atoms are near each other and close to a conducting plate.
These are reminiscent of effects seen between two macroscopic objects and
a wall \cite{zaheer,rodriguez}; our work generalized that given in 
Ref.~\cite{lopez}.  We broke up the three-body terms into three- and 
four-scattering contributions; although both are comparable at short distances,
as expected, the former dominate for atoms far from the plate.
In order to work out the three-body effects for an atom
interacting with two half-planes, constituting a slit in a conducting
plane,  or more generally, facing wedges, in Sec.~\ref{sec:twedge} we work
out the scattering matrix for a single wedge, and then, 
in Sec.~\ref{sec:atom-wedge}, apply that to
recalculate the two-body repulsion found in Ref.~\cite{Milton:2011ni}. 
Because of the complexity of the calculation, a scalar analog is also
considered in the Appendix. 
We then go on to consider three-body effects between a polarizable atom and a
pair of parallel conducting cylinders in Sec.~\ref{sec:2cyl}.  
In Ref.~\cite{Milton:2011ed}  we showed that for an anisotropic atom
moving along a line perpendicular to but not intersecting a perfectly
conducting cylinder, and polarizable along that same line, a repulsive
force occurs near the cylinder, provided the distance of closest approach
is sufficiently large compared to the radius of the cylinder.  (The same
does not occur for a sphere.)  Here we consider the three-body effects
due to a second cylinder parallel to the first, so the pair forms an aperture,
perpendicular to which the anisotropic atom moves.  We adapt our formalism
to this case, where a multipole expansion is also possible, and show that
when the distance between the cylinders is sufficiently large, repulsion is
not affected  by the three-body corrections, since the latter are very small.
Finally, in Sec.~\ref{sec:anisotropic}, we present a calculation that suggests
that highly anisotropic atoms, necessary to exhibit the repulsive effects
we are considering, may be beyond reach.  Therefore, as in the numerical
calculations of Ref.~\cite{Levin:2010zz}, 
it may be more appropriate to
consider the interaction with highly anisotropic conducting nanoparticles, such
as needles, as suggested in related work on negative Casimir-Polder entropy
\cite{Milton:2014tca}.

\section{Three-body Casimir Energy}\label{sec:3body}
The multiple scattering formulation has proved exceptionally useful in 
computing Casimir energies for complex configurations.  It is usually 
presented in terms of
potentials, where the potential stands in for the deviation of the permittivity
from its vacuum value, for instance.  Here, however, we wish from the outset to
consider perfect conductors, so we give the formulation entirely in 
terms of scattering matrices.  In particular, we wish to analyze 
three-body effects.  The formalism we apply here appears in many places;
recent examples are Refs.~\cite{torque2,Milton:2014tca}.  We use natural units,
with $\hbar=c=1$, and Heaviside-Lorentz electromagnetic units, except for
the definition of the polarizability. 

The quantum vacuum energy, with the bulk vacuum energy subtracted, 
is in general given by
\be
E=\frac{i}2\Tr \ln \bm{\Gamma}\bm{\Gamma}_0^{-1},
\ee
where the $\Tr$ symbol represents a trace over tensor indices as well as
spatial coordinates.  Here $\bm{\Gamma}_0$ is the free Green's dyadic,
which, for a given frequency $\omega$, can be written as 
\be
\bm{\Gamma}_0(\mathbf{r,r'})=(\bm{1}\omega^2+\bm{\nabla}
\bm{\nabla})G_0(|\mathbf{r-r'}),\label{fgd}
\ee
in terms of the free Helmholtz Green's function
\be
G_0(R)=\frac{e^{i|\omega| R}}{4\pi R},\quad R=|\mathbf{r-r'}|.
\label{fhelmfg}
\ee
The full Green's dyadic $\bm{\Gamma}$ satisfies 
 the same differential equation as the free Green's dyadic,
\be
\bm{\Gamma}_0^{-1}\bm{\Gamma}=\bm{1},
\ee
where 
\be
\bm{\Gamma}_0^{-1}=\frac1{\omega^2}\bm{\nabla}\times\bm{\nabla}\times -{\bm 1}
=\frac1{\omega^2}\left[\bm{\nabla\nabla}-(\nabla^2+\omega^2)\bm{1}\right].
\label{greensop}
\ee
Here we have adopted a matrix notation for both the tensor indices and the 
spatial coordinates, so
\be
\bm{1}=\bm{1} \delta(\mathbf{r-r'}),
\ee
where on the right $\bm{1}$ refers to the tensor indices only.  The 
conducting surfaces $S$
appear through boundary conditions on the Green's dyadic,
\be
\mathbf{\hat n}\times\bm{\Gamma}\Big|_S=0,\label{pcbc}
\ee
where $\mathbf{\hat n}$ is the outward normal to the surface at the 
point in question.
We may define the scattering matrix $\mathbf{T}$ by
\be
\bm{\Gamma}=\bm{\Gamma}_0+\bm{\Gamma}_0\mathbf{T}\bm{\Gamma}_0,
\ee
so that
\be
\mathbf{T}=-\bm{\Gamma}_0^{-1}+\bm{\Gamma}_0^{-1}\bm{\Gamma}\bm{\Gamma}_0^{-1}.
\label{teematrix}
\ee

Now we turn to the quantum interaction of three bodies.  It seems easiest to
start with the situation where the bodies may be described by potentials 
$\mathbf{V}_i$, $i=1$, 2, 3, and then write the result in a form in which 
only the $\mathbf{T}$ operators appear, so it applies to the conducting 
boundary problem, defined by
Eq.~(\ref{pcbc}).  The total potential is 
$\mathbf{V}=\mathbf{V}_1+\mathbf{V}_2+\mathbf{V}_3$, 
and the vacuum energy is given by the trace-log of 
\be
\bm{\Gamma}\bm{\Gamma}_0^{-1}=(\bm{1}-\bm{\Gamma}_0\mathbf{V})^{-1}, 
\ee
or
\be 
E=-\frac{i}2\Tr\ln(\bm{1}-\bm{\Gamma}_0\mathbf{V}).
\ee
Now it is easy to see that
\bea
&&1-\bm{\Gamma}_0(\mathbf{V}_1+\mathbf{V}_2+\mathbf{V}_3)
=(\bm{1}-\bm{\Gamma}_0\mathbf{V}_1-\bm{\Gamma}_0\mathbf{V}_2)\nn\\
&&\quad\times\left[\bm{1}-(\bm{1}-\bm{\Gamma}_1\mathbf{V}_2)^{-1}\bm{\Gamma}_1\mathbf{V}_2
\bm{\Gamma}_1\mathbf{V}_3(\bm{1}-\bm{\Gamma}_1\mathbf{V}_3)^{-1}\right]\nn\\
&&\qquad\times(\bm{1}-\bm{\Gamma}_0\mathbf{V}_1)^{-1}
(\bm{1}-\bm{\Gamma}_0\mathbf{V}_1-\bm{\Gamma}_0\mathbf{V}_3).\label{interform}
\eea
Here we have introduced the Green's dyadic belonging to potential $i$ alone,
\be
\bm{\Gamma}_i=(\bm{1}-\bm{\Gamma}_0\mathbf{V}_i)^{-1}\bm{\Gamma}_0.
\ee
Now in Eq.~(\ref{interform}) the factors before and after the square
brackets  refer to only
one- and two-body interactions (the latter referring to interactions between
bodies 1 and 2, and 1 and 3, respectively), so the two-body interactions 
between 2 and 3, and three-body interactions are all contained in the quantity 
in square brackets.  Now, in terms of the potential, the corresponding 
scattering matrix is
\be
\mathbf{T}_i =\mathbf{V}_i(\bm{1}-\bm{\Gamma}_0\mathbf{V}_i)^{-1}.
\ee
Introducing the modified scattering matrix defined by Shajesh and
Schaden \cite{shajesh}, 
\be
\mathbf{\tilde T}=\mathbf{T}\bm{\Gamma}_0,
\ee
and using the cyclic property of the trace, we
find the two and three body terms:
\be
E_{23}=-\frac{i}2\Tr\ln(\bm{1}-
\mathbf{\tilde T}_2\mathbf{\tilde T}_3),\label{e23}
\ee
which is sometimes called the $TGTG$ formula, and
\bea
E_{123}&=&-\frac{i}2\Tr\ln\bigg(\bm{1}-\mathbf{X}_{23}\bigg[\mathbf{X}_{21}
\mathbf{\tilde T}_2(\bm{1}+\mathbf{\tilde T}_1)
\mathbf{X}_{31}\nn\\
&&\times\mathbf{\tilde T}_3
(\bm{1}+\mathbf{\tilde T}_1)-\mathbf{\tilde T}_2\mathbf{\tilde T}_3\bigg]
\bigg),\label{3-body}
\eea
where
\be
\mathbf{X}_{ij}=(\bm{1}-\mathbf{\tilde T}_i\mathbf{\tilde T}_j)^{-1}.
\ee
Expression (\ref{3-body}) 
has a rather evident geometrical interpretation in terms of
multiple scattering off the three objects.
This is not written in as symmetrical a form as in Ref.~\cite{shajesh}, but
is somewhat simpler, particular for the Casimir-Polder applications that 
follow, where body 1 represents the atom, so is treated weakly.

\section{Polarizable atoms between parallel conducting plates}
\label{sec:parallel}
\begin{figure}
\includegraphics[scale=.8]{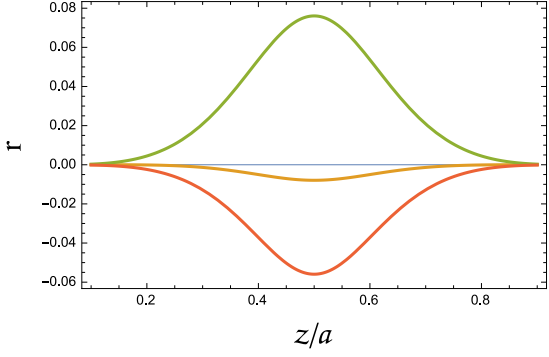}
\caption{\label{figppatom} Three-body contributions to the
Casimir-Polder interaction of an anisotropically polarizable atom between
two parallel polarizable plates. What is plotted is the ratio
 of the
three-body contribution relative to the total energy,
$r=E_{123}/E_{CP}$.
 The three-body contributions are generally very small.  They become
appreciable only far from both plates, where the Casimir-Polder energy
is very small.  Plotted is the ratio for isotropic atoms (middle curve),
atoms polarizable only in the direction perpendicular to the plates (top
curve), and polarizable only parallel to the plates (bottom curve).  In
the  case when only $\alpha_{zz}\ne0$, the sign of the three-body correction is
the same as that of the two-body term.}
\end{figure}

As a simple check of the machinery developed in the previous section, we
revisit the calculation of the interaction energy of
an anisotropically polarizable atom between parallel conducting
plates, a geometry first analyzed by Barton \cite{barton}.  Since we know
the Green's dyadic $\bm{\Gamma}$
for parallel plates, it is easy to derive the interaction
energy from the general Casimir-Polder formula
\be 
E_{CP}=-\int_{-\infty}^\infty \,d\zeta\Tr \bm{\alpha}\cdot\bm{\Gamma},
\ee
where the integration is over imaginary frequency, 
$\omega\to i\zeta$.\footnote{This replacement requires knowledge of the
analytic properties of the integrand.  There can be serious subtleties
involved in this ``Euclidean transformation,'' for example, see
Ref.~\cite{guerout14}.  See also Ref.~\cite{berman14}.}
Here the (in general, anisotropic) polarizability of the atom is
$\bm{\alpha}$.  In the following we will assume $\bm{\alpha}$ is independent
of frequency, i.e., we are working in the static approximation.
The interaction energy for one conducting
plate at $z=0$, one at $z=a$, and the atom at $z=Z$, $0<Z<a$, is
\bea
E_{CP}&=&\frac{\alpha_{11}+\alpha_{22}-\alpha_{33}}{4\pi a^4}\zeta(4)\nn\\
&&\quad\mbox{}-\frac{\tr\bm{\alpha}}
{8\pi a^4}\left[\zeta(4,Z/a)+\zeta(4,1-Z/a)\right],
\eea
in terms of the Hurwitz zeta function.\footnote{The particular combinations of 
Hurwitz zeta functions occurring here and in the following are striking. 
Such combinations occur in several places, for instance  when considering 
the Casimir energy for two 
parallel plates in $D$ dimensional spacetimes, where the argument 
4 is replaced by $D$.  Two  of the first papers in this 
direction  are Refs.~\cite{alnes06,alnes07}.  See also Ref.~\cite{brevik08}.
Related structures appear for the theory of the piecewise uniform
string \cite{li91,brevik12,brevik03}.} 
Here the two-body interactions between the atom and one or the other plate
are isolated by extracting the parts singular as $Z\to 0$ or $Z\to a$:
\bea
\zeta(4,Z/a)&=&\left(\frac{a}Z\right)^4+\zeta(4,1+Z/a),\nn\\
\zeta(4,1-Z/a)&=&\left(\frac{a}{a-Z}\right)^4+\zeta(4,2-Z/a).
\eea
The total Casimir-Polder energy is the sum of two-body and three-body
terms,
\be
E_{CP}=E_{12}+E_{13}+E_{123},
\ee
where 1 denotes the atom, 2 the lower plate, and 3 the upper plate.
Here
\begin{subequations}
\be
E_{12}= -\frac{\tr\bm{\alpha}}{8\pi Z^4},\quad 
E_{13}= -\frac{\tr\bm{\alpha}}{8\pi (a-Z)^4},\ee
and
\bea
E_{123}&=&\frac{\alpha_{11}+\alpha_{22}-\alpha_{33}}{4\pi a^4}\zeta(4)\nn\\
&&\quad\mbox{}-\frac{\tr\bm{\alpha}}
{8\pi a^4}\left[\zeta(4,1+Z/a)+\zeta(4,2-Z/a)\right].\nn\\
\label{e123}
\eea
\end{subequations}
Note that the first term in $E_{123}$ is independent of $Z$, so it does not contribute
to the Casimir-Polder force on the atom, but is a Casimir-Polder correction
to the Casimir force between the plates.
The two-body energies overwhelmingly dominate the Casimir-Polder interaction,
as shown in Fig.~\ref{figppatom}.  For isotropic atoms, the largest three
body correction is only a 0.8\% reduction
at the midpoint between the plates, where the
energy is very small.   For atoms only polarizable parallel to 
the $z$ direction the three-body correction is an 8\% increase 
at the place where the energy is the smallest,
while for purely transversely polarizable atoms, the three-body correction
is a 6\% reduction at the midpoint.  

\subsection{Multiple-scattering calculation}\label{section:mspara}
\label{ssec:msp}
Since in more general situations we do not have an exact solution available,
we want to calculate the three-body corrections (\ref{e123}),
using the multiple-scattering
formula (\ref{3-body}).  For this purpose, we need to compute the scattering
operators for the three bodies.

The scattering matrix for the atom is simply
\be
\mathbf{T}_1(\mathbf{r,r'})=\mathbf{V}_1(\mathbf{r,r'})
=4\pi\bm{\alpha}\delta(\mathbf{r-R})
\delta(\mathbf{r-r'}),
\ee
where $\mathbf{R}=(0,0,Z)$ is the position of the atom.
The free electromagnetic Green's dyadic can be written as
\be
\bm{\Gamma}_0(\mathbf{r-r'})=\int\frac{(d\mathbf{k}_\perp)}{(2\pi)^2}
e^{i\mathbf{k_\perp\cdot(r-r')_\perp}}\bm{\gamma}_0(z,z'),
\ee
where
\be
\bm{\gamma}_0(z,z')
=(\mathbf{E+H})\frac1{2\kappa}e^{-\kappa|z-z'|},
\ee
with the usual abbreviation $\kappa=\sqrt{k^2+\zeta^2}$.  Here $\mathbf{E}$
and $\mathbf{H}$ are matrices corresponding to the transverse electric (TE)
and transverse magnetic (TM) modes,
\bea
\mathbf{E}&=&-\zeta^2\left(\begin{array}{ccc}
s^2&-cs&0\\
-cs&c^2&0\\
0&0&0\end{array}\right),\nn\\
\mathbf{H}(z,z')&=&\left(\begin{array}{ccc}
c^2\partial_z\partial_{z'}&cs\partial_z\partial_{z'}&ikc\partial_z\\
cs\partial_z\partial_{z'}&s^2\partial_z\partial_{z'}&iks\partial_z\\
-ikc\partial_{z'}&-iks\partial_{z'}&k^2\end{array}\right).\label{ehforpp}
\eea
Here $k^2=\mathbf{k}_\perp^2$ and $c$ ($s$) is the cosine 
(sine) of the angle between
the direction of $\mathbf{k}_\perp$ and the $x$-axis, $c=k_x/k$, $s=k_y/k$.
The polarization operators are transverse, in the sense that
\be
i\mathbf{k}_\perp\cdot \mathbf{H}+\partial_z\mathbf{\hat z}\cdot\mathbf{H}=0,
\ee
and similarly for $\mathbf{E}$.
Thus the modified scattering matrix for the atom is
\bea
\mathbf{\tilde T}_1(\mathbf{r,r}')
&=&4\pi\bm{\alpha}\delta(z-Z)\delta(\mathbf{r_\perp})
\int\frac{(d\mathbf{k_\perp})}{(2\pi)^2}e^{i\mathbf{k_\perp\cdot(r-r')_\perp}}
\nn\\
&&\quad\times(\mathbf{E+H})(z,z')\frac1{2\kappa}e^{-\kappa|z-z'|}.
\eea

The following composition
properties of the $\mathbf{E}$ and $\mathbf{H}$ operators are easily checked:
\begin{subequations}
\label{ehprops}
\bea
\mathbf{E H}&=&0,\\
\mathbf{E}\mathbf{E}&=&-\zeta^2\mathbf{E},\\
\mathbf{H}(z,z')\mathbf{H}(z'',z''')&=&(k^2+\partial_{z'}\partial_{z''})
\mathbf{H}(z,z''').\nn\\
\label{hprop}
\eea
\end{subequations}

For a single plate, say a conducting plate 2 at $z=0$, we have the
reduced Green's dyadic in the form
\be
\bm{\gamma}=\mathbf{E}g^E+\mathbf{H}g^H,
\ee
where
\bea
g^{E,H}(z,z')&=&g_0(z,z')\mp\frac1{2\kappa}e^{-\kappa(|z|+|z'|)}
\left\{\begin{array}{c}
1,\\
\mbox{sgn}(z)\mbox{sgn}(z'),\end{array}\right.\nn\\
 g_0(z,z')&=&\frac1{2\kappa}e^{-\kappa|z-z'|}.
\eea
Then the reduced modified scattering matrix is
\be
\mathbf{\tilde t}_2(z,z')=\bm{\gamma}_0^{-1}
(\bm{\gamma}-\bm{\gamma}_0)(z,z').
\ee
This is evaluated by using the transverse property of $\mathbf{E}$ and
$\mathbf{H}$ [hence, the vector Helmholtz operator
reduces to $-\nabla^2+\zeta^2$---see Eq.~(\ref{greensop})] and
\begin{subequations}\label{disc}
\bea
\left(-\frac{d^2}{dz^2}+\kappa^2\right)e^{-\kappa|z|}&=&2\kappa\delta(z),\\
\left(-\frac{d^2}{dz^2}+\kappa^2\right)\mbox{sgn}(z)e^{-\kappa|z|}&=&
2\delta'(z)e^{-\kappa |z|}.
\label{disc2}
\eea
\end{subequations}
Thus the modified scattering matrix for a conducting plate at $z=0$ is
\bea
\mathbf{\tilde{t}}_2(z,z')&=&\frac1{\zeta^2}
\bigg[\mathbf{E}(z,z')\delta(z) e^{-\kappa|z'|}
\nn\\
&&\quad\mbox{}+\mathbf{H}(z,z')\frac1\kappa\delta'(z)
e^{-\kappa|z|}\mbox{sgn}(z') e^{-\kappa|z'|}\bigg].\nn\\
\eea
However, because $\delta'(z)$ is an instruction to integrate by parts and 
evaluate at $z=0$, which action is on the exponential propagators occurring 
in every case,  we can use, as in Ref.~\cite{Milton:2014tca},
\be
\mathbf{\tilde t}_2(z,z')=\frac1{\zeta^2}\mathbf{(E-H)}(z,z')
\delta(z)e^{-\kappa|z'|}.
\ee

Then it is easy to see that the two-body interaction energy between the
atom and the plate is as expected:
\be 
E_{12}=\frac{i}2\Tr \mathbf{\tilde T}_1\mathbf{\tilde T}_2
=-\frac{\tr \bm{\alpha}}{8\pi Z^4}.
\ee

\subsection{CP interaction between atom and two parallel conducting plates}

The three-body interaction is worked out by simplifying the multiple-scattering
formula (\ref{3-body}) for the case when
there is only one interaction with the atom, since that coupling is always
weak:
\bea
E_{123}&=&\frac{i}2\Tr \mathbf{X}_{23}
\big(\mathbf{\tilde T}_2\mathbf{\tilde T}_1
\mathbf{\tilde T}_2\mathbf{\tilde T}_3+\mathbf{\tilde T}_2\mathbf{\tilde T}_3
\mathbf{\tilde T}_1\mathbf{\tilde T}_3\nn\\
&&\quad\mbox{}+\mathbf{\tilde T}_2\mathbf{\tilde T}_1
\mathbf{\tilde T}_3+\mathbf{\tilde T}_2\mathbf{\tilde T}_3\mathbf{\tilde T}_1
\big),\label{e123a}
\eea
where the $\mathbf{\tilde T}$ operators are given in Sec.~\ref{section:mspara}.

Let us look at the $\mathbf{E}$ and $\mathbf{H}$ parts separately. For the
TE part,
\be
\mathbf{X}_{23}\mathbf{E}=\mathbf{E}\frac1{1-e^{-2\kappa a}}\delta(z),
\ee
so
\bea
E_{123}^{\rm TE}&=&\frac1{4\pi^2}\int d\zeta\,(d\mathbf{k}_\perp)
\tr\bm{\alpha}
\left(-\frac{\mathbf{E}}{\zeta^2}\right)
\frac{e^{-2\kappa a}}{1-e^{-2\kappa a}}\left(-\frac{\zeta^2}{2\kappa}\right)
\nn\\
&&\quad\times\left[e^{-2\kappa Z}+e^{-2\kappa(a-Z)}-2\right], 
\eea
where integrating over the directions of $\mathbf{k}_\perp$ gives for the
trace
\be
\tr\bm{\alpha}
\left(-\frac{\mathbf{E}}{\zeta^2}\right)\to\frac12(\alpha_{11}+\alpha_{22}).
\ee
Thus the TE contribution is
\bea
E_{123}^{\rm TE}&=&-\frac{\alpha_{11}+\alpha_{22}}{12\pi}\int_0^\infty
d\kappa\,\kappa^3\frac1{e^{2\kappa a}-1}\nn\\
&&\quad\times\left[-2+e^{-2\kappa Z}
+e^{-2\kappa(a-Z)}\right]\nn\\
&=&-\frac{\alpha_{11}+\alpha_{22}}{32\pi a^4}
\bigg[-2\zeta(4)+\zeta(4,1+Z/a)\nn\\
&&\qquad\mbox{}+\zeta(4,2-Z/a)\bigg].\label{te123}
\eea

The TM contribution is similarly worked out, with the result
\bea
E_{123}^{\rm TM}&=&-\frac1{2\pi}\int_0^\infty \frac{d\kappa\,\kappa^3}
{e^{2\kappa a}-1}\bigg\{\frac{\alpha_{11}+\alpha_{22}}2\nn\\
&&\quad\times\left[e^{-2\kappa Z}
+e^{2\kappa(a-Z)}-2\right]\nn\\
&&\quad\qquad\mbox{}+\frac23\alpha_{33}\left[e^{-2\kappa Z}+
e^{-2\kappa(a-Z)}+2\right]\bigg\}\nn\\
&=&-\frac{3(\alpha_{11}+\alpha_{22})}{32\pi a^4}\nn\\
&&\quad\times\left[-2\zeta(4)
+\zeta(4,1+Z/a)+\zeta(4,2-Z/a)\right]\nn\\
&&\mbox{}-\frac{\alpha_{33}}{8\pi a^4}
\left[2\zeta(4)+\zeta(4,1+Z/a)+\zeta(4,2-Z/a)\right].\nn\\
\eea
Adding this to the TE contribution (\ref{te123}), gives the three-body energy 
(\ref{e123}).

\begin{figure}
\includegraphics[scale=.8]{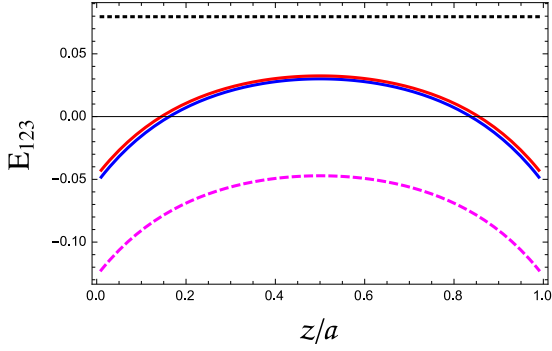}
\caption{\label{fig3and4} Three-body contribution to the
Casimir-Polder interaction of an isotropically polarizabable atom between
two parallel polarizable plates. The interaction energy is given in units
of $\alpha/a^4$.  The upper dotted horizontal line is the
three-scattering approximation, the lower dashed curve is the four-scattering
approximation, the upper solid
(red) curve is the sum of these two contributions, 
which is only slightly above the the full three-body energy (solid blue
curve).} 
\end{figure}

The three-body corrections are dominated by the three- and four-scattering
contributions, given by the explicit scattering terms in Eq.~(\ref{e123a}), 
with
the multiple-reflection quantity $\mathbf{X}_{23}$ 
set equal to 1.  That translates
into replacing the zeta functions in Eq.~(\ref{e123}) by their leading terms,
$\zeta(4)\to1$, $\zeta(4,x)\to 1/x^4$.  Figure \ref{fig3and4} compares the 
exact three-body corrections to the leading three- and four-scattering
approximations.  (It is geometrically obvious why the odd-scattering terms
give contributions which are independent of the position of the atom, because
the path length is then an integer multiple of the plate separation.)

\section{Casimir-Polder Interaction between Atom and Wedge}
\label{sec:atomwedge}

\begin{figure}
\begin{center}
 \includegraphics{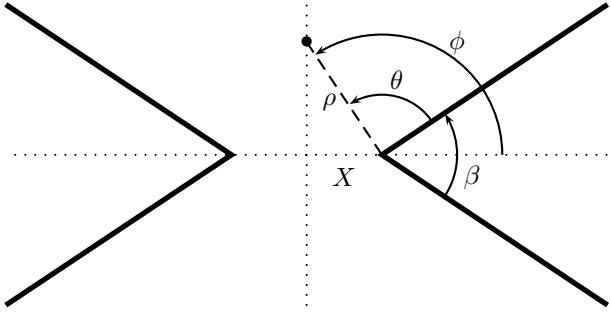}
\end{center}
\caption{\label{fig:fw}
Two facing conducting wedges, with an anisotropically polarizable atom
passing along a line perpendicular to  the symmetry plane of the wedges.}
\end{figure}

Our goal had been  to compute the three-body corrections
for an atom near an aperture created by two facing wedges, as
shown in Fig.~\ref{fig:fw}.
Here we have two parallel conducting wedges, with opening angles $\beta$, 
whose apexes are separated
by a distance $2X$.  As both interior wedge angles go to zero, the situation 
reduces to two conducting half-planes, lying in the same plane, 
with a gap between them.
The Casimir-Polder interaction between the two wedges
and an anisotropically polarizable molecule, located at coordinates $\rho$, 
$\phi$ relative to the apex of one wedge, is to be computed.  In particular, 
we wish to 
study the three-body interaction, which involves scattering off all three
objects, as a correction to the more elementary calculation of the interaction
between the atom and a single wedge, which is given in 
Ref.~\cite{Milton:2011ni}.  For the latter, repulsion can be achieved 
for the angle $\phi$ sufficiently close to $\pi$, provided
$\beta$ is smaller than 108$^\circ=1.88$ radians.

\subsection{Scattering Matrix for a Perfectly Conducting Wedge}
\label{sec:twedge}

In Ref.~\cite{Milton:2011ni} we gave the Green's dyadic for a single  perfectly
conducting wedge, in terms of polar coordinates based at a point on the apex of
the wedge.  We write this most conveniently in terms of the quantity $\kappa=
\sqrt{k^2+\zeta^2}$, where $\zeta=-i\omega$ is the imaginary frequency and the
wavenumber in the longitudinal direction or $z$ direction, 
the direction perpendicular to the plane
of the figure, is $k$.  With $\theta$ defined as the angle from the top surface
of the wedge, that Green's dyadic is (the prime on the summation sign is
an instruction to count the $m=0$ term with half weight)
\bea
&&\bm{\Gamma}(\mathbf{r,r'})=\frac{2p}\pi\sum_{m=0}^\infty{}'
\int_{-\infty}^\infty\frac{dk}{2\pi}\nn\\
&&\quad\times\bigg[\mathbf{E}(\mathbf{r,r'})\cos mp\theta\cos mp\theta'\nn\\
&&\qquad\mbox{}+\mathbf{H}(\mathbf{r,r'})
\sin mp\theta\sin mp\theta'\bigg]e^{ik(z-z')}\nn\\
&&\qquad\times\frac{I_{mp}(\kappa\rho_<)K_{mp}(\kappa \rho_>)}{-\kappa^2}.
\label{wedgegf}
\eea
Here $\rho_>$ ($\rho_<$) is the greater (lesser) of $\rho$, $\rho'$, and
$p=\pi/(2\pi-\beta)$. The electric and magnetic polarization
dyadic operators are
\begin{subequations}
\label{eandh}
 \bea
\mathbf{E(r,r')}&=&
-\left(\bm{\hat\rho}\frac1\rho\partial_\theta-\bm{\hat\theta}
\partial_\rho\right)\left(\bm{\hat\rho}'\frac1{\rho'}\partial_{\theta'}
-\bm{\hat\theta}'
\partial_{\rho'}\right)\nn\\
&&\qquad\times(\nabla^2_\perp-k^2)\nn\\
&=&-\zeta^2(\mathbf{\hat z}\times
\bm{\nabla_\perp})
(\mathbf{\hat z}\times\bm{\nabla}'_\perp)\nn\\
&=&-\nabla^2(\mathbf{\hat z}
\times\bm{\nabla}_\perp)(\mathbf{\hat z}\times\bm{\nabla}'_\perp),
\label{defewedge}\\
\mathbf{H(r,r')}&=&\left[ik\left(\bm{\hat\rho}\partial_\rho+\bm{\hat\theta}
\frac1\rho\partial_\theta
\right)-\mathbf{\hat z}\nabla_\perp^2\right]\nn\\
&&\quad\times\left[-ik\left(\bm{\hat\rho}'
\partial_{\rho'}
+\bm{\hat\theta}'\frac1{\rho'}\partial_{\theta'}\right)-\mathbf{\hat z}
\nabla_\perp^{\prime2}\right]\nn\\
&=&\left(ik\bm{\nabla}_\perp-\mathbf{\hat z}\kappa^2\right)
\left(-ik\bm{\nabla}'_\perp-\mathbf{\hat z}\kappa^2\right)\nn\\
&=&[\bm{\nabla}\times(\bm{\nabla}\times\mathbf{\hat z})]
[\bm{\nabla}'\times(\bm{\nabla}'\times\mathbf{\hat z})].
\eea
\end{subequations}
[Here the polarization operators differ from those given previously
in Eq~(\ref{ehforpp}) by the replacements $\mathbf{E}\to -\nabla^2_\perp 
\mathbf{E}$, $\mathbf{H}\to -\nabla^2_\perp \mathbf{H}$.  
For a further discussion of the
properties of the polarization operators, see Ref.~\cite{torque2}.]
In the second forms in Eq.~(\ref{eandh}) we have used the modified Bessel
 equation, that
is, that for either modified Bessel function
\be
(-\nabla_\perp^2+\kappa^2) e^{i\nu(\theta-\theta')} 
\left\{\begin{array}{c}I_\nu(\kappa\rho)\\K_\nu(\kappa\rho)\end{array}\right.
=0.
\ee

In the following we will need the composition properties of these operators, 
analogous to those in Eq.~(\ref{ehprops}):
\begin{subequations}
 \bea
\mathbf{E(r,r')H(r'',r''')}&=&\mathbf{H(r,r')E(r'',r''')}=0.\\
\mathbf{E(r,r')E(r'',r''')}&=&\mathbf{E(r,r''')}\nabla^{\prime2}_\perp
(\nabla_\perp^{\prime\prime2}-k^2)\nn\\
&\to& \kappa^2\zeta^2\mathbf{E(r,r''')},\\
\mathbf{H(r,r')H(r'',r''')}&=&\mathbf{H(r,r''')}
\nabla^{\prime2}_\perp(\nabla_\perp^{\prime\prime2}-k^2)\nn\\
&\to& \kappa^2\zeta^2\mathbf{H(r,r''')},
\eea
\end{subequations}
where we will understand that after differentiation, the intermediate 
coordinates $\mathbf{r'}$ and $\mathbf{r''}$ are set equal and integrated
over; that is, a spatial matrix multiplication is understood.

To construct the modified scattering matrices, we need the free Green's dyadic,
which we can write as Eq.~(\ref{fgd}),
where a representation for the scalar Helmholtz Green's function 
in cylindrical coordinates is
\bea
G_0(\mathbf{r-r'})&=&\int_{-\infty}^\infty
\frac{dk}{2\pi}e^{ik(z-z')}\frac1{2\pi}\sum_{m=-\infty}^\infty e^{im(\theta-
\theta')}\nn\\ 
&&\quad\times I_m(\kappa\rho_<)K_m(\kappa\rho_>).
\eea
It is easy to verify that in terms of the mode operators,
\be
\mathbf{\Gamma}_0(\mathbf{r,r'})=-\frac1{\nabla_\perp^2}(\mathbf{E+H)(r,r')} 
G_0(\mathbf{r-r'}).
\ee

Using the above, we find the modified scattering matrix for the atom to be
\bea
\mathbf{\tilde T}_{\rm atom}(\mathbf{r,r'})&=&
4\pi\bm{\alpha}\delta(\mathbf{r-R})
(\mathbf{E+H})(\mathbf{r,r'})\int\frac{dk}{2\pi}e^{ik(z-z')}\nn\\
&&\,\,\times\frac1{2\pi}
\sum_{m=-\infty}^\infty e^{im(\theta-\theta')}\frac{I_m(\kappa\rho_<)
K_m(\kappa\rho_>)}{-\kappa^2}.
\nn\\
\label{tatom}
\eea

To work out the $\mathbf{\tilde T}$ matrix for the wedge, we start from 
Eq.~(\ref{teematrix}),
\be
\mathbf{\tilde T}=\bm{\Gamma}_0^{-1}(\bm{\Gamma}-\bm{\Gamma}_0).
\label{ttwiddle}
\ee
The inverse free Green's function is the differential operator given in
Eq.~(\ref{greensop}). It is easy to check that
\begin{subequations}
\label{gm1eh}
\bea
\bm{\Gamma}_0^{-1}\mathbf{E}&=&\mathbf{E}\frac{\kappa^2-\nabla_\perp^2}{\omega^2},\\
\bm{\Gamma}_0^{-1}\mathbf{H}&=&\mathbf{H}\frac{\kappa^2-\nabla_\perp^2}{\omega^2}.
\eea
\end{subequations}
The Helmholtz operator appearing as the last factor here would 
annihilate the scalar
Green's functions appearing in Eq.~(\ref{wedgegf}), except on the boundaries, 
where
the normal derivatives give contributions to the scattering matrix that live
entirely on the surface of the wedge.  
This is precisely the same as what occurred for the planes
in Sec.~\ref{ssec:msp}: see Eq.~(\ref{disc}).  Here, because we are considering
the region exterior to the wedge, we interpret the angular mode functions as 
follows:
($\theta\in[0,\Omega]$)
\begin{subequations}
\bea
\cos mp\theta&\to&\cos mp\theta\,\eta(\theta)\,\eta(\Omega-\theta),\\
\sin mp\theta&\to&\sin mp\theta\,\eta(\theta)\,\eta(\Omega-\theta),
\eea
\end{subequations}
where $\Omega=2\pi-\beta$ is the exterior wedge angle, and the step function is
defined by
\be
\eta(x) =\left\{\begin{array}{cc}
                 1,&x>0,\\0, &x<0.
                \end{array}\right.\ee
Then we see that
 \bea
[\partial_\theta^2+(mp)^2]\cos mp\theta&=&\delta'(\theta)
-(-1)^m\delta'(\theta-\Omega),\nn\\{}
[\partial_\theta^2+(mp)^2]\sin mp\theta&=&mp[\delta(\theta)
-(-1)^m\delta(\theta-\Omega)].\nn\\
\eea
From this we can immediately read off the modified
scattering matrix for the wedge:
\bea
&&\mathbf{\tilde T}_{\rm wedge}(\mathbf{r,r'})
=-\frac{2p}\pi\sum_{m=0}^\infty{}'
\int_{-\infty}^\infty
\frac{dk}{2\pi}\frac1{\zeta^2}
\nn\\
&&\times\bigg\{\mathbf{E(r,r')}\left[\delta'(\phi-\beta/2)-(-1)^m
\delta'(\phi+\beta/2)\right]\nn\\
&&\qquad\quad\times\cos mp(\phi'-\beta/2)\nn\\
&&\quad\quad\mbox{}+mp\mathbf{H(r,r')}\left[\delta(\phi-\beta/2)
-(-1)^m\delta(\phi+\beta/2)\right]\nn\\
&&\qquad\quad\times\sin mp(\phi'-\beta/2)\bigg\}e^{ik(z-z')}\nn\\
&&\quad\times\frac1{\kappa^2\rho^2}I_{mp}(\kappa\rho_<)K_{mp}
(\kappa\rho_>),\label{twedge}
\eea
where we have shifted to the angular variable $\phi$ 
measured from the symmetry plane of the wedge,
as shown in Fig.~\ref{fig:fw}, $\phi$, $\phi'\in [\beta/2,2\pi-\beta/2]$, 
and the delta
functions are understood to be periodically extended, with period $2\pi$.

\subsection{Two-body Calculation}
\label{sec:atom-wedge}
We now want to use this multiple scattering formalism, 
particularly Eq.~(\ref{e23}), to reproduce the results found in Ref.~\cite{Milton:2011ni}.
Putting together the scattering matrix for the atom, Eq.~(\ref{tatom}), and
that for the wedge, Eq.~(\ref{twedge}), we obtain the following expression
for the Casimir-Polder energy,
\bea
E_{aw}&=&-i\frac{p}{4\pi^3}\tr\bm{\alpha}\int \frac{d\zeta dk}{\kappa^2}
\int_0^\infty \frac{d\rho'}{\rho'}
\sum_{m=-\infty}^\infty\sum_{m'=-\infty}^\infty\nn\\
&&\quad\times \left[m\mathbf{E(r,r'')}-m'p
\mathbf{H(r,r'')}\right]e^{im'p(\phi''-\beta/2)}\nn\\
&&\quad\times 
e^{im\phi}\left[e^{-im\beta/2}
-(-1)^{m'}e^{im\beta/2}\right]\nn\\
&&\quad\times I_m(\kappa\rho_<)K_m(\kappa\rho_>)
I_{|m'|p}(\kappa\tilde\rho_<)K_{|m'|p}(\kappa\tilde\rho_>).\nn\\
\label{twobody}
\eea
Here $\rho_{<,>}$ is the lesser, greater of $\rho, \rho'$, and
$\tilde\rho_{<,>}$ is the lesser, greater of $\rho'',\rho'$.
After the differentiations contained in $\mathbf{E}$ and $\mathbf{H}$ are
performed, the coordinates $\rho''$ and $\phi''$ are to be set
equal to  $\rho$ and $\phi$, respectively.

Although we can carry out the $m$ summation, or the $\rho'$ integration,
it seems difficult to bring Eq.~(\ref{twobody}) into the closed form given
in Ref.~\cite{Milton:2011ni}.  So we initially will
 content ourselves with a special case,
$\beta=\pi$ or $p=1$, that is, the interaction of an atom with an infinite
conducting plane.  In that case, we may as well ultimately set $\phi=\phi''
=\pi$.  Then only Bessel functions of integer order occur, and both the 
$m$ and $m'$
sums can be carried out, using the addition theorem,
\be
K_0(\kappa P)=\sum_{m=-\infty}^\infty e^{im(\phi-\phi')}I_m(\kappa\rho_<)K_m(\kappa\rho_>),
\ee
where
\be
P=\sqrt{\rho^2+\rho^{\prime2}-2\rho\rho'\cos(\phi-\phi')}.
\ee
Then the energy can be written as
\bea
E_{ap}&=&-\frac{i}{2\pi^2}\int_0^\infty \frac{d\kappa}\kappa \int_0^\infty 
\frac{d \rho'}{\rho'}\mbox{tr}\,\bm{\alpha}\mathbf{(E-H)(r,r''})\nn\\
&&\quad\times\bigg[K_0(\kappa\sqrt{\rho^2+\rho^{\prime2}-2\rho\rho'\sin\phi})
\nn\\
&&\quad\times\frac1i\frac\partial{\partial\phi''}
K_0(\kappa\sqrt{\rho^{\prime\prime2}
+\rho^{\prime2}-2\rho''\rho'\sin\phi''})\nn\\
&&\quad\mbox{}-(\sin\phi\to-\sin\phi)\bigg]\bigg|_{\rho''\to\rho,\,\phi''\to
\phi=\pi}
\nn\\
&=& \frac1{2\pi^2}\int_0^\infty d\kappa \int_0^\infty d\rho\,
\mbox{tr}\,\bm{\alpha}(\mathbf{E-H)(r,r'')}\nn\\
&&\quad\times\bigg[\frac{\rho''\cos\phi''}{\sqrt{
\rho^{\prime2}+\rho^{\prime\prime2}-2\rho'\rho''\sin\phi''}}\nn\\
&&\quad\times K_0'(\kappa(\sqrt{\rho^{\prime2}+
\rho^{\prime\prime2}-2\rho'\rho''\sin\phi''})\nn\\
&&\qquad\times K_0(\kappa\sqrt{\rho^{2}
+\rho^{\prime2}-2\rho\rho'\sin\phi})\nn\\
&&\quad\mbox{}+(\sin\phi\to-\sin\phi)\bigg]
\bigg|_{\rho''\to\rho,\phi''\to\phi=\pi}.
\eea

The simplest situation occurs when the atom is only polarizable along the
axis of the wedge,
 $\bm{\alpha}=\mathbf{\hat z\hat z}\alpha_{zz}$.  Then 
$\mbox{tr}\,\bm{\alpha}\mathbf{E}=0$, 
$\mbox{tr}\,\bm{\alpha}\mathbf{H}=\kappa^4\alpha_{zz}$, and we have
\bea E_{ap}&=&
-\frac{\alpha_{zz}}{\pi^2}\int_0^\infty d\rho'\frac \rho{\sqrt{\rho^2
+\rho^{\prime2}}}\int_0^\infty d\kappa\,\kappa^4 \nn\\
&&\quad\times K_0(\kappa \sqrt{
\rho^2+\rho^{\prime2}})
K_1(\kappa \sqrt{\rho^2+\rho^{\prime2}})\nn\\
&=&-\frac{\alpha_{zz}}{8\pi \rho^4},
\eea
which is the expected Casimir-Polder energy.
With only a bit more effort, we find the familiar
result for arbitrary polarization,
\be
E_{\rm CP}=-\frac{\mbox{tr}\,\bm{\alpha}}{8\pi\rho^4}.
\ee

Given the difficulty of even analytically reproducing the two-body
correction given in Ref.~\cite{Milton:2011ni}, it is not surprising
that we did not get very far with the three-body calculation.  In the Appendix
we consider the scalar analog for the two-body effect, and although we get
a bit further, we have been unable to reproduce the analytic result obtained
by the direct calculation.\footnote{However, it
is possible to recast the $\tilde T_1\tilde T_2$ expression into
the form of  the direct CP energy $E_{CP}=
\Tr\int d\zeta\alpha(G-G_0)$---see the Appendix.}
  So we turn, instead, to another problem, the
interaction between an atom and a pair of cylinders.

\section{Casimir-Polder Interaction of Atom with Two Parallel Cylinders}
\label{sec:2cyl}
\begin{figure}
\centering
\includegraphics{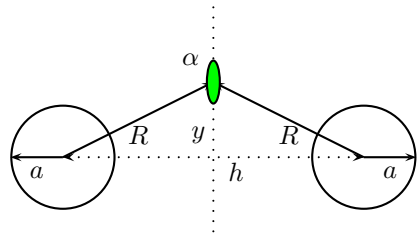}
\caption{\label{fig2cyl}
An anisotropically polarizable atom, denoted by $\alpha$, is 
symmetrically located relative to two identical parallel perfectly conducting
cylinders (with axes coming out of the page).  
The centers of the cylinders are separated by a
distance $h$, and each has radius $a$. The atom is on the line bisecting the
line connecting the centers of the two cylinders, but a distance $y$ above it.
The atom is a distance $R$ from the center of either cylinder.  The angle
$\theta$ is defined by $\cos\theta=y/R$.}
\end{figure}

The difficulties in extracting usable expressions for three-body effects
for the atom-wedge-wedge problem has
to do with the lack of a scale, so multipole expansions, for example, are
not applicable.  Therefore, we turn to another example, that of an atom
interacting with a pair of parallel cylinders, illustrated in 
Fig.~\ref{fig2cyl}.
Here it is assumed the two cylinders are identical, with radius $a$, and
their centers are separated by a distance $h$.  The atom is located on the
line bisecting the line connecting the centers of the cylinders, a distance
$R$ from the center of each, and a height $y$ above the centerline.  We will
consider the case when the atom is only polarizable along the bisecting
line.

\subsection{Scattering matrix for cylinder}
We first need to compute the scattering matrix for the cylinder.  The
Green's dyadic, given in Ref.~\cite{DeRaad:1981hb}, part of which
appears in Ref.~\cite{Milton:2011ed}, 
can be written, with slight notational changes,  as\footnote{As noted in
Ref.~\cite{CaveroPelaez:2004xp}, only the terms in the Green's functions 
involving modified Bessel functions, and not powers of the radial coordinates, 
contribute to the electric and magnetic fields.} 
\bea
\bm{\Gamma}_{\rm cyl}(\mathbf{r,r'})&=&
-\sum_{m=-\infty}^\infty \int\frac{dk}{(2\pi)^2} e^{im(
\phi-\phi')}e^{ik(z-z')}\nn\\
&&\quad\times\left[\mathbf{E}F_m(\rho,\rho')
+\mathbf{H}G_m(\rho,\rho')\right],
\eea
where the TE and TM Green's functions are, outside the cylinder,
\begin{subequations}
\bea
F_m(\rho,\rho')&=&\frac1{\kappa^2}\bigg[I_m(\kappa \rho_<)K_m(\kappa \rho_>)
\nn\\
&&\quad\mbox{}-\frac{I_m'(\kappa a)}{K_m'(\kappa a)}
K_m(\kappa \rho)K_m(\kappa \rho')\bigg],
\\
G_m(\rho,\rho')&=&\frac1{\kappa^2}\bigg[I_m(\kappa \rho_<)K_m(\kappa \rho_>)
\nn\\
&&\quad\mbox{}-\frac{I_m(\kappa a)}{K_m(\kappa a)}
K_m(\kappa \rho)K_m(\kappa \rho')\bigg],
\eea
\end{subequations}
and the polarization operators are the same as given in Eqs.~(\ref{eandh}).
The modified scattering matrix is given by Eq.~(\ref{ttwiddle}), where,
because of the tranversality of the polarization operators, the inverse
free Green's operator may be replaced by $\bm{\Gamma}_0^{-1}\to\frac1{\zeta^2}
(\nabla^2-\zeta^2)$ [cf.~Eq.~(\ref{gm1eh})], 
which vanishes everywhere but on the surface of the
cylinder.  Because we have a perfectly conducting body, $F_m(\rho,\rho')=0$ if
$\rho'>a>\rho$, and so for the TE function with $\rho'>a$
\begin{subequations}
\bea
&&(F_m - F_m^{0})(\rho,\rho')=-\frac1{\kappa^2}K_m(\kappa \rho')\bigg[
\frac{I_m'(\kappa a)}{K_m'(\kappa a)}\nn\\
&&\qquad\times K_m(\kappa \rho)\eta(\rho-a)
+I_m(\kappa \rho)\eta(a-\rho)\bigg].
\eea
Similarly, the TM functions are
\bea
&&(G_m - G_m^{0})(\rho,\rho')=-\frac1{\kappa^2}K_m(\kappa \rho')\bigg[
\frac{I_m(\kappa a)}{K_m(\kappa a)}\nn\\
&&\qquad\times K_m(\kappa \rho)\eta(\rho-a)
+I_m(\kappa \rho)\eta(a-\rho)\bigg].
\eea
\end{subequations}
Then a simple calculation leads to the scattering matrix
\bea
\tilde{\mathbf{T}}_{\rm cyl}&=&\sum_{m=-\infty}^\infty \int\frac{dk}{(2\pi)^2} 
e^{ik(z-z')}e^{im(\phi-\phi')}\frac1{\kappa^2\zeta^2 a}\nn\\
&&\quad\times\bigg[\mathbf{E}
\frac1\kappa \frac1\rho\frac{\partial}{\partial \rho}\rho\delta(\rho-a)
\frac{K_m(\kappa \rho')}{K_m'(\kappa a)}\nn\\
&&\qquad\mbox{}-\mathbf{H}\delta(\rho-a)\frac{K_m(\kappa \rho')}
{K_m(\kappa a)}\bigg].
\eea

To check its validity, we reproduce the two-body interaction between
one cylinder and the anisotropic atom, for which we obtain
\bea
E_{12}&=&-\frac12\int_{-\infty}^\infty \frac{d\zeta}{2\pi}\Tr 
\tilde{\mathbf{T}}_{\rm atom}\tilde{\mathbf{T}}_{\rm cyl}\nn\\
&=&-\int\frac{d\zeta d k}{(2\pi)^2}\sum_{m=-\infty}^\infty\tr\frac{\bm{\alpha}}
{\kappa^2}\bigg[\frac{I_m'(\kappa a)}{K_m'(\kappa a)}\mathbf{E}(\mathbf{r,r'})
\nn\\
&&\qquad\mbox{}+\frac{I_m(\kappa a)}{K_m(\kappa a)}
\mathbf{H}(\mathbf{r,r'})\bigg]\nn\\
&&\qquad\times K_m(\kappa \rho)K_m(\kappa \rho')\bigg|_{\mathbf{r=r'=R}}.
\eea
This is the general result, which may be derived by simpler means.  In
particular, for the situation envisaged in Fig.~\ref{fig2cyl}, where
$R=h/(2\sin\theta)$, and the atom is only polarizable along the $y$ direction,
we obtain the formulas given in Ref.~\cite{Milton:2011ed}\footnote{The signs
of the energies given there, and in Fig.~3 of that reference, should
be reversed. All physical conclusions in that paper, however, are correct.}
in terms of the distance of closest approach $R_0=h/2$,
\begin{subequations}
\bea
E^{\rm TM}_{\rm CP}&=&-\frac\alpha{4\pi}\frac{\sin^4\theta}{R_0^4}
\sum_{m=-\infty}^\infty \int_0^\infty dx \, x\frac{I_m(\kappa a\sin\theta/R_0)}
{K_m(\kappa a\sin\theta/R_0)}\nn\\
&&\quad\times\left[m^2 K_m^2(x)\sin^2\theta+x^2
K_m^{\prime2}(x)\cos^2\theta\right],\nn\\
\\
E^{\rm TE}_{\rm CP}&=&\frac\alpha{4\pi}\frac{\sin^4\theta}{R_0^4}
\sum_{m=-\infty}^\infty \int_0^\infty dx \,x\frac{I'_m(\kappa a\sin\theta/R_0)}
{K'_m(\kappa a\sin\theta/R_0)}\nn\\
&&\quad\times\left[m^2 K_m^2(x)\cos^2\theta+x^2
K_m^{\prime2}(x)\sin^2\theta\right].\nn\\
\eea
\end{subequations}
These formulas show that repulsion indeed occurs along the bisector 
($y$) direction
provided the distance of closest approach $R_0$ is larger than about 7 times
the radius of the cylinder; in this case the $m=0$ term dominates, and the
TM contribution is much larger than the always attractive TE contribution, as
we will see below.

\subsection{Three-body correction}
We now want to see if the above repulsive effect survives when the effect of
both bodies are included.  By virtue of the symmetry seen in 
Fig.~\ref{fig2cyl}, the two-body forces in the $y$-direction
 are doubled.  That is, if the atom is called body 1, and
the cylinders are 2 and 3, respectively, the two-body terms are just
\be
E_{\rm 2-body}=E_{12}+E_{13}.
\ee
The three-body terms are computed from Eq.~(\ref{e123a}).  In view of the 
remarks above, because we are considering large distances between the 
cylinders, it should suffice to consider the three- and four-scattering 
terms, the higher terms being suppressed,
\bea
&&E_{\rm 3-body}\approx E_{123}+E_{132}+E_{1232}+E_{1323},\nn\\  &&E_{123}
=-\frac12\Tr\tilde{\mathbf{T}}_1\tilde{\mathbf{T}}_2\tilde{\mathbf{T}}_3,
\quad
E_{1232}=-\frac12\Tr\tilde{\mathbf{T}}_1\tilde{\mathbf{T}}_2
\tilde{\mathbf{T}}_3\tilde{\mathbf{T}}_2.\nn\\
\eea
Further, we expect dominance by the $m=0$ TM mode.  However, we can effect
considerable simplification before we make that last approximation.  Indeed,
the formula for the TM mode 3-scattering energy  simplifies to
\bea
E^{\rm TM}_{123}
&=&\int\frac{d\zeta\,dk}{(2\pi)^2}\frac1{\kappa^2}\tr\bm{\alpha}
\mathbf{H}(\mathbf{r,\tilde r})\nn\\
&&\quad\times\sum_{mm'}
\int_0^{2\pi}\frac{d\tilde\phi'}{2\pi} e^{i(m\phi-m'\tilde\phi)}
 e^{i(m'\tilde\phi'-m\phi')}\nn\\
&&\quad\times\frac{I_m(\kappa a)}{K_m(\kappa a)}
\frac{K_m(\kappa  \rho')}{K_{m'}(\kappa a)}
K_m(\kappa\rho)K_{m'}(\kappa\tilde\rho)
\bigg|_{\rho=\tilde{\rho}=R},\nn\\
\label{g123}
\eea
and the corresponding TE 3-scattering energy is
\bea
E^{\rm TE}_{123}
&=&\int\frac{d\zeta\,dk}{(2\pi)^2}\frac1{\kappa^2}\tr\bm{\alpha}
\mathbf{E}(\mathbf{r,\tilde r})\nn\\
&&\quad\times\sum_{mm'}
\int_0^{2\pi}\frac{d\tilde\phi'}{2\pi} e^{i(m\phi-m'\tilde\phi)}
 e^{i(m'\tilde\phi'-m\phi')}\nn\\
&&\quad\times\frac{I'_m(\kappa a)}{K'_m(\kappa a)}
\frac{K'_m(\kappa  \rho')}{K'_{m'}(\kappa a)}\frac{d\rho'}{d a}
K_m(\kappa\rho)K_{m'}(\kappa\tilde\rho)
\bigg|_{\rho=\tilde{\rho}=R},\nn\\
\label{g123e}
\eea
where $a$ and $\tilde\phi'$ are the cylindrical 
coordinates of a point on the surface of the second cylinder relative to
the central axis of that cylinder, the same point being located at cylindrical
coordinates $\rho'$ and $\phi'$ relative to the central
axis of the first cylinder. These coordinates are related by
\be
\rho^{\prime2}=h^2+a^2-2ah\cos\tilde\phi',\quad \tan\phi'
=\frac{a\sin\tilde\phi'}{a\cos\tilde\phi'-h}.
\ee
The atom is located in the two coordinate systems at
\be
\mathbf{R}=(R,\phi,0)=(R,\tilde \phi,0),
\ee
where 
\be
\phi=\frac\pi2+\theta,\quad \tilde\phi=\frac\pi2-\theta.
\ee
It would now be straightforward to work out the multipole expansion of 
Eqs.~(\ref{g123}), (\ref{g123e})
that is, a power series expansion in powers of $a/h$.  We will content 
ourselves with the lowest term, which means we can set $m=0$,
because only for small values of $a/h$ do we have 2-body repulsion. 
Higher terms are suppressed by powers of $a/h$.  In this limit,
the TE energy is completely negligible, because of the appearance 
of derivatives of Bessel functions.  The behavior of the Bessel functions 
for small argument makes this point clear:
\begin{subequations}
\bea
I_0(z)&\sim& 1+\frac14 z^2,\quad K_0(z)\sim-\gamma-\ln\frac{z}2,\\
I'_0(z)&\sim&\frac12z,\quad K_0'(z)\sim-\frac1z,\quad z\to0.
\eea
\end{subequations}
 Then we 
get a very simple explicit formula, again for an atom polarizable only in the
bisector ($y$) direction,
\bea
E_{123}^{\rm TM}&\sim& \frac{\alpha_{yy}}{4\pi R_0^4}\cos^2\theta\sin^4\theta
\nn\\
&&\times\int_0^\infty dx \,x^3\frac{K_0(2x  \sin\theta)I_0(x a \sin\theta/R_0)}
{K_0^2(x a \sin\theta/R_0)}K_0^{\prime2}(x).\nn\\
\eea
Here $h=2R_0$ is the separation distance between the axes of the cylinders.
In the same approximation, the 4-scattering contribution is also given simply:
\bea
E_{1232}^{\rm TM}&\sim& -\frac{\alpha_{yy}}{4\pi R_0^4}\cos^2\theta\sin^4\theta
\nn\\
&&\times\int_0^\infty dx \,x^3\frac{K^2_0(2x  \sin\theta)
I_0(x a \sin\theta/R_0)}
{K_0^3(x a \sin\theta/R_0)}K_0^{\prime2}(x).\nn\\
\eea

\begin{figure}
\centering
\includegraphics[scale=0.6]{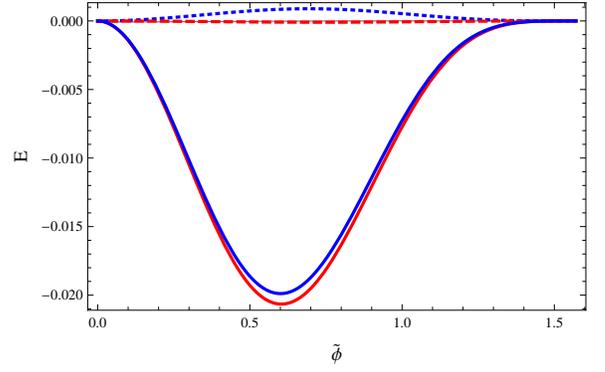}
\caption{\label{2and3fig}
The Casimir-Polder energy of an anisotropic atom or nanoparticle
passing on the symmetry line perpendicular to a pair of identical perfectly
conducting parallel cylinders.  The energy, apart from a factor of $\alpha/(
4\pi R_0^4)$, is plotted versus $\tilde\phi=\pi/2-\theta$, that is,
the angle between the line connecting the axes of the cylinders and the line
connecting the atom with the center of either cylinder.  The 
bottom curve (red) is the two-body energy, 
the top curve (dotted-blue) is the three-scattering
correction, the second-from-top curve (dashed red) is the four-scattering 
term, while the second  curve from the bottom (blue) is the total 
Casimir-Polder energy.
The energies are plotted for $a/R_0=0.01$. Because
$a/R_0\ll1$  it is sufficient to include only $m=0$ for the three-body
corrections. Also, the TE contributions  are completely negligible.
 It is seen that the three-body effects, in fact, are very small,
and do not significantly alter  the repulsion
 between the atom and the pair of cylinders, and the four-scattering
terms are quite negligible.}
\end{figure}

These TM corrections are plotted in Fig.~\ref{2and3fig}  
as a function of $\tilde\phi=\pi/2-\theta$, so $\tilde\phi=0$
at the position of the atom closest to the cylinders, and compared to the
two body contributions. (All the TE corrections are completely negligible.)
The figure shows the two-body CP energy, entirely dominated by the $m=0$ TM
contribution, and the $m=0$ three-body correction, dominated by the 
three-scattering terms, in the limit of large
distance between the cylinders. The TE contributions were computed, but found,
as expected, to be completely insignificant in the large distance regime.
 The TM three-body correction is not 
quite negligible, but does not affect the Casimir-Polder repulsion discovered 
in Ref.~\cite{Milton:2011ed}.

\section{Difficulty of achieving high atomic anisotropy}\label{sec:anisotropic}

In order for Casimir--Polder repulsion to be possible, the atom interacting 
with a body must have a sufficiently anisotropic polarizability tensor. 
Defining an anisotropy factor $\gamma$ according to 
\be
\bdalpha = \alpha_{zz} \uvz\uvz + \gamma\alpha_{zz} \uvx\uvx+\gamma\alpha_{zz}
\uvy\uvy,
\ee
implying
\be
\gamma = \frac{\mathrm{Tr}(\bdalpha)-\alpha_{zz}}{2\alpha_{zz}},
\ee
we found in Ref.~\cite{Milton:2011ni} that for an
anisotropic atom interacting with a half-plane,
 the critical value of $\gamma$ was $1/4$. Values $\gamma<1/4$ give repulsion 
in certain circumstances, whereas for $\gamma>1/4$ no repulsion is possible.
The non-retarded interaction of an atom and a circular aperture was considered 
in Ref.~\cite{eberlein11} and the critical value of $\gamma$ was found to be 
$1/4$ also in this case \cite{Milton:2011ni}.

In this section we investigate what minimal value of the anisotropy parameter 
can be achieved by preparing an atom in an excited eigenstate $|nlm\rangle$. 
Here, $n$ denotes the principal quantum number,
$l=0,\ldots n-1$ is the quantum number for the orbital angular
momentum and $m=-l\ldots l$ that for its $z$-component.
The question is of great interest especially in the light of recent advances 
in experimental techniques using Rydberg atoms, 
atoms excited to high principal quantum numbers, near boundaries 
\cite{kubler10,crosse10}, 
and noting that Rydberg atoms can take highly anisotropic shapes.

Because of the close spacing of energy levels for highly excited states, and 
the fact that transitions to the states nearest in energy to $|nlm\rangle$ 
dominate the CP energy \cite{crosse10}, the Casimir--Polder interaction of a 
Rydberg molecule is essentially non-retarded even at atomically 
large separations, up to hundreds of micrometers. 
It was shown that in such cases the interaction energy 
is proportional to the atomic dipole moment tensor 
\cite{ellingsen10,ellingsen11},
\be
E_{nlm} =-\langle \mathbf{d}\mathbf{d}\rangle:\bm{\nabla\nabla}G
\bigg|_{\omega=0}
=-\langle \mathbf{d}\mathbf{d}\rangle:\boldsymbol{\Gamma}_{\omega=0},
\ee
according to Eq.~(\ref{fgd}).
For convenience we will work with the ratio
\be
  q=\langle{d}^2_{zz}\rangle/\langle{\mathbf{d}}^2\rangle.
\ee
Defining $\gamma$ similarly as before
\be
  \gamma = \frac{\langle{\mathbf{d}}^2\rangle-\langle{d}^2_{zz}\rangle}
{2\langle{d}^2_{zz}\rangle} = 
\frac{1}{2}\left(\frac1{q}-1\right),
\ee
we will consider energy eigenstates $|nlm\rangle$ such that anisotropy becomes
 maximal, i.e., 
$\gamma$ becomes minimal and $q$ maximal.

To evaluate the anisotropy parameter, we insert the completeness
relation $\sum_{n'l'm'}|n'l'm'\rangle\langle n'l'm'|={I}$,
\begin{equation}
\label{eq2}
q=\frac{\sum_{n'l'm'}\langle nlm|{d}_z|n'l'm'\rangle
 \langle n'l'm'|{d}_z|nlm\rangle}
 {\sum_{n'l'm'}\langle nlm|{\mathbf{d}}|n'l'm'\rangle\sprod
 \langle n'l'm'|{\mathbf{d}}|nlm\rangle}
\end{equation}
The dipole-matrix elements can conveniently be calculated by means of
the Wigner--Eckart theorem \cite{0982,0981}
\begin{equation}
\label{eq3}
\langle n'l'm'|{d}_s|nlm\rangle
 =(-1)^{l'-m'}\begin{pmatrix}l'&1&l\\-m'&s&m\end{pmatrix}
 \langle n'l'||{\mathbf{d}}||n l\rangle\;,
\end{equation}
where $\langle n'l'||{\vect{d}}||n l\rangle$ denotes the reduced
matrix element and the Wigner 3-$j$ symbol can be given in terms of
Clebsch--Gordan coefficients as \cite{1005}
\begin{equation}
\label{eq4}
\begin{pmatrix}j_1&j_2&j\\m_1&m_2&m\end{pmatrix}
= \frac{(-1)^{j_1-j_2-m}}{\sqrt{2j+1}}
\langle j_1m_1j_2m_2|j\,-\!m\rangle\;.
\end{equation}
Substituting these relations into Eq.~(\ref{eq2}) and using the
orthonormality relation \cite{1005}
\begin{equation}
\label{eq5}
\sqrt{2j+1}\sum_{m_1m_2}
\begin{pmatrix}j_1&j_2&j\\m_1&m_2&m\end{pmatrix}
\begin{pmatrix}j_1&j_2&j'\\m_1&m_2&m'\end{pmatrix}
=\delta_{jj'}\delta_{mm'}\;,
\end{equation}
we find
\begin{equation}
\label{eq6}
q\to q_{lm}
=\sum_{l'}|\langle l\, m\, 1\,0|l'm\rangle|^2\;.
\end{equation}
As expected from the symmetry of the problem, the anisotropy
parameter depends neither on the reduced matrix element nor on the
principal quantum numbers.

The Clebsch-Gordan coefficients in Eq.~(\ref{eq6}) can be evaluated
explicitly, leading to \cite{1005}
\begin{equation}
\label{eq7}
q_{lm}
=\frac{l^2-m^2}{(2l-1)(2l+1)}
 +\frac{(l+1)^2-m^2}{(2l+1)(2l+3)}\;.
\end{equation}
For a given $l$, the anisotropy parameter obviously takes its maximum
value for $m=0$,
\begin{equation}
\label{eq8}
q_{l0}
=\frac{l^2}{(2l-1)(2l+1)}
 +\frac{(l+1)^2}{(2l+1)(2l+3)}
=\frac{2l(l+1)-1}{4l(l+1)-3}\;.
\end{equation}
The latter expression is equal to $1/3$ for $l=0$, approaches $1/2$
for $l\to\infty$ and takes it maximum value
\begin{equation}
\label{eq9}
q_{10}=\tfrac{3}{5} ~~ \Longrightarrow ~~ \gamma = \tfrac{1}{3}
\end{equation}
for $l=1$. The maximally anisotropic eigenstate of orbital angular
momentum is thus a $p$ state.

Since $q_{lm}$ is positive for any given choice of quantum numbers,
it immediately follows that the anisotropy 
parameter $\gamma$
is bounded 
below by $1/3$
for any incoherent superposition of energy eigenstates.
It is possible that stronger anisotropies could be realizable with a coherent 
superposition of states.  However, the more likely venue for discovering such
repulsive effects would be with anisotropic particles, such as elongated 
needles.

\section{Conclusions}
One of the principal features of Casimir or quantum vacuum forces is that they
are not additive. Unlike classical electrodynamics, one cannot simply sum
pairwise forces.  Such approximations clearly are invalid even for the simplest
situations of parallel plates.  This, of course, makes calculations more
challenging.

In this paper, we have explored some aspects of three-body interactions in
the context of Casimir-Polder forces between an anisotropically polarizable
atom or nanoparticle and two conducting surfaces.  First we examined the role
of such forces involving an atom between two perfectly conducting plates,
a well-known problem \cite{barton}, but one in which we can test our
formalism and isolate explicitly the three-body terms.  Then we turned to
the interaction of such an atom with a pair of wedges; we reproduced the
repulsive effects seen for an atom interacting with a single wedge
\cite{Milton:2011ni}, but
a closed form for the three-body correction remains elusive.  So we then
examined the interaction of an anisotropic atom with a pair of parallel
cylinders.  The two-body repulsive effect found earlier \cite{Milton:2011ed}
was reproduced, and the three-body correction was computed in the limit
of large separation between the cylinders, which is the regime where repulsion
is expected.  The three-body correction is non-negligible in this limit,
is completely captured by the TM three-scattering approximation, but
is too small to affect  the earlier-found repulsion.

\acknowledgments
KAM, EKA, and PP  thank  the US National Science Foundation, the US Department 
of Energy, the Simons Foundation, and the Julian Schwinger Foundation
for partial support of this research. Some of this work was accomplished while
KAM was partially supported by Laboratoire Kastler Brossel, CNRS; he also
thanks NTNU for hospitality.
We also thank K. V. Shajesh and M. Schaden for useful discussions.
SYB gratefully acknowledges support by the German Research Council
(grant BU 1803/3-1) and the Freiburg Institute for Advanced Studies.

\appendix*

\section{Scalar analog for atom-wedge problem}
\label{sec:scalar}
In this Appendix we will  consider a scalar analog of the atom-wedge problem.
Let the atom be described by the potential
\be V_{\rm atom}=4\pi\alpha\delta(\mathbf{r-R}),
\ee
where $\mathbf{R}=(R,\phi,0)$ is the position of the atom.  
The modified scattering matrix for the (Dirichlet) wedge is
[note that the sign is reversed compared to Eq.~(\ref{ttwiddle})]
\be \tilde T_{\rm wedge}=1-G_0^{-1} G_w,\ee
where
\bea
G_w(\mathbf{r,r'})&=&
\frac{2p}\pi\int_{-\infty}^\infty\frac{dk}{2\pi}
e^{ik(z-z')}\sum_{m=1}^\infty \sin mp(\phi-\beta/2)\nn\\
&&\quad\times\sin mp(\phi'-\beta/2)
 I_{mp}(\kappa\rho_<)K_{mp}(\kappa\rho_>).\nn\\
\eea
Applying $G_0^{-1}=-\nabla^2+\zeta^2$ we obtain the scattering matrix
on the wedge:
\bea
\tilde T_w&=&\frac{2p}\pi
\int_{-\infty}^\infty \frac{dk}{2\pi}\sum_{m=1}^\infty
e^{ik(z-z')}\frac{mp}{\rho^2}[\delta(\phi-\beta/2)\nn\\
&&\quad\mbox{}-(-1)^m\delta(\phi+\beta/2)]
\sin mp(\phi'-\beta/2)\nn\\
&&\quad\times I_{mp}(\kappa\rho_<)K_{mp}(\kappa\rho_>).
\eea%

The two-body energy is given by
\be
E_{12}=\frac{i}2\int\frac{d\omega}{2\pi}\Tr\tilde T_1\tilde T_2=-\frac12\int_{-
\infty}^\infty \frac{d\zeta}{2\pi}\Tr V_1G_0\tilde T_2.
\ee
If we use the two-dimensional representation for the free propagator,
\be
G_0(\mathbf{r})=\int_{-\infty}^\infty \frac{dk_z}{2\pi} e^{i k_z z}
\frac1{2\pi}K_0(\kappa|\mathbf{r}_\perp|),\ee
the two-body energy can be written as
\bea
E_{12}&=&-\frac\alpha{\pi^2}p^2\int_0^\infty d\kappa\,\kappa\int_0^\infty
\frac{d\rho'}{\rho'}\sum_{m=1}^\infty m\sin mp(\phi-\beta/2)\nn\\
&&\quad\times I_{mp}(\kappa\rho_<)K_{mp}(\kappa\rho_>)\nn\\
&&\quad\times[K_0(\kappa P_+)-(-1)^m K_0(\kappa P_-)],\label{e12sc}
\eea
where $P_\pm$ is the distance between the atom and a point on the upper (lower)
wedge boundary,
\be
P_{\pm}=\sqrt{R^2+\rho^{\prime2}-2R\rho'\cos(\phi\mp\beta/2)}.
\ee
Now the integral of the three Bessel functions can be performed:
\bea
&&\int_0^\infty dt\,t \,I_\nu(\xi t) K_\nu(t) K_0(P_{\pm} t/\rho_>)\nn\\
&=&\frac1{2\xi \sin(\phi-\beta/2)}\sum_{n=0}^\infty \xi^{\nu+n+1}
\frac{\sin(n+1)(\phi-\beta/2)}{\nu+n+1},\nn\\
\eea
where $\xi=\rho_</\rho_>$.  The radial integrals are then easy, and we are
immediately led to
\bea
E_{12}&=&-\frac{\alpha p^2}{\pi^2 R^2}\sum_{m=1}^\infty m\sin mp(\phi-\beta/2)
\nn\\
&&\quad\times\sum_{n=0}^{\infty}\frac1{(n+mp)(n+mp+2)}\nn\\
&&\quad\times\bigg[\frac{\sin(n+1)(\phi-\beta/2)}
{\sin(\phi-\beta/2)}\nn\\
&&\qquad\mbox{}-(-1)^m\frac{\sin(n+1)(\phi+\beta/2)}{\sin(\phi+\beta/2)}
\bigg].
\eea
Now we replace the sum over $m$ by an integral,
\bea
\sum_{m=1}^\infty \frac1{mp+N} e^{imp\theta}&=&\sum_{m=1}^\infty \int_0^\infty
dt\,e^{-t(mp+N)}e^{imp\theta}\nn\\
&=& \int_0^\infty dt\, e^{-tN}\frac1{e^{p(t-i\theta)}-1}.
\eea
Then the $n$ sum can be carried out as a geometric series, and the result is
a single integral,
\bea
&&E_{12}=-\frac{\alpha p^2}{4\pi^2 R^2}\int_0^\infty dt\sinh t\sinh pt\sin(\phi
-\beta/2)\nn\\
&&\times\bigg\{\frac1{[\cosh pt-\cos p(\phi-\beta/2)]^2}\frac1{\cosh t
-\cos(\phi-\beta/2)}\nn\\
&&\mbox{} +\frac1{[\cosh pt+\cos p(\phi-\beta/2)]^2}\frac1{\cosh t
-\cos(\phi+\beta/2)}\bigg\}.\nn\\
\label{ms2body}
\eea 

Alternatively, we can directly calculate the two-body energy from
\be
E_{12}=\frac12\int_{-\infty}^\infty \frac{d\zeta}{2\pi} V_1(G-G_0).\label{alt}
\ee
which may be directly evaluated in closed form:
\be
E_{12}=-\frac\alpha{8\pi R^2}\left[\frac{p^2}{\sin^2p(\phi-\beta/2)}+\frac13
(1-p^2)\right].\label{scalarcp}
\ee
The form (\ref{ms2body}) may be evaluated straightforwardly in analytic
form for the two special cases $p=1$ ($\beta=\pi$, that is, a plane), and
$p=1/2$ ($\beta=0$, that is, a half-plane), in agreement with 
Eq.~(\ref{scalarcp}),
\begin{subequations}
\bea
E_{12}(p=1)&=&-\frac\alpha{8\pi R^2}\frac1{\sin^2\theta},\\
E_{12}(p=1/2)&=&-\frac\alpha{32\pi R^2}\left(\frac1{\sin^2\theta/2}+1\right).
\eea
\end{subequations}
For other values of $p$ the analytic evaluation of Eq.~(\ref{ms2body}) seems
nontrivial; however, the integral is rapidly convergent, and the numerical
coincidence with the closed form (\ref{scalarcp}) is easily verified. 

Not surprisingly, it is possible to show that the explicit 
form of Eq.~(\ref{alt})
follows from the multiple scattering form of the two-body energy (\ref{e12sc}),
written in terms of a sum over four Bessel functions and a sum over $m$ and
$m'$ as in Eq.~(\ref{twobody}).  This involves using the integral [notation
as in Eq.~(\ref{twobody}), with $\nu=m'p$],
\bea
&&\int\frac{d\rho'}{\rho'}I_m(\kappa\rho_<)K_m(\kappa\rho_>)I_\nu(\kappa\tilde
\rho_<)K_\nu(\kappa\tilde\rho_>)\nn\\
&=&\frac1{m^2-\nu^2}[K_\nu(\kappa R)I_\nu(\kappa R)-I_m(\kappa R)
K_m(\kappa R)],\nn\\
\eea
where the resulting two terms in the energy are summed over $m$ and $m'$,
respectively, using
\begin{subequations}
\bea
\sum_{m=-\infty}^\infty \frac{e^{im\theta}}{m^2-\nu^2}&=&-\frac\pi\nu
\frac{\cos\nu(\theta-\pi)}{\sin\pi\nu},\\
\sum_{m'=1}^\infty\frac{m'p\sin m'p}{m^2-(m'p)^2}&=&-\frac\pi{2p}
\sin\frac{m}p(\pi-p\theta)\csc\frac{m\pi}p,\nn\\
\\
\sum_{m'=1}^\infty(-1)^{m'}\frac{m'p\sin m'p}{m^2-(m'p)^2}&=&-\frac\pi{2p}
\sin m\theta\csc\frac{m\pi}p.
\eea
\end{subequations}

\end{document}